\documentclass{article}
\usepackage{arxiv}

\usepackage{amsmath}
\usepackage{amssymb}
\usepackage[pdftex]{graphicx}
\usepackage{siunitx}
\graphicspath{{./figures/}}

\newcommand{\wb}{\mathbf{w}}
\newcommand{\bb}{\mathbf{b}}
\newcommand{\xb}{\mathbf{x}}
\newcommand{\iu}{\mathsf{i}}
\newcommand{\ju}{\mathsf{j}}
\newcommand{\Wn}{W^{+}}
\newcommand{\Xn}{X^{+}}
\newcommand{\Bn}{B^{+}}
\newcommand{\Cn}{\mathbb{C}^{+}}
\newcommand{\Wa}{W^{\mathrm{a}}}
\newcommand{\Ba}{B^{\mathrm{a}}}
\newcommand{\Xa}{X^{\mathrm{a}}}
\newcommand{\dphi}{d\varphi}
\newcommand{\dphiW}{\dphi_{\Wn}}

\newcommand{\dphiC}{\dphi_{\Cn}}

\newcommand{\phia}{\phi^{\mathrm{a}}}

\title{The InSAR absolute phase amid singularities}

\author{S. Zwieback \vspace{2mm} \\
	Geophysical Institute \& Department of Geosciences\\
	University of Alaska Fairbanks\\
	Fairbanks, AK, USA \\
	\texttt{szwieback@alaska.edu}
}
\date{December 26, 2025}

\begin{document}
\maketitle

\begin{abstract}
The radar interferometric absolute phase is considered essential for estimating topography and displacements. However, the conventional definition---that absolute phase is proportional to the range difference---cannot be applied to general targets. Here, a universal observational definition is proposed, which is easiest to describe for differential interferometry: The absolute phase is determined by temporally unwrapping the phase as the intermediate secondary acquisition time sweeps from the primary to the secondary acquisition time. This absolute phase is typically not directly observable because a continuous series of observations is required. The absolute phase of a point target is proportional to the range difference, matching the conventional definition. For general targets undergoing cyclic change (returning to their initial state), the absolute phase may be nonzero and cannot be interpreted as a range difference. When a phase singularity (vanishing coherence) occurs at an intermediate time, the absolute phase becomes undefined, a situation termed an absolute phase singularity. Absolute phase singularities complicate absolute phase reconstruction through multifrequency techniques and through unwrapping multidimensional interferograms. They leave no trace in an interferogram, but unwrapping paths need to avoid those across which the absolute phase jumps by nonzero integer multiples of $2 \pi$. Mathematical analyses identify conditions for unwrapping-based reconstruction up to a constant, accounting for absolute phase singularities, undersampling and noise. The general definition of the absolute phase and the mathematical analyses enable a comprehensive appraisal of InSAR processing chains and support the interpretation of observations whenever low coherence engenders phase and absolute phase singularities.
\end{abstract}

\section{Introduction}
In radar interferometry, the absolute phase is an essential but nevertheless ill-defined concept \cite{bamler98,rosen00}. According to Madsen et al. \cite{madsen93}, the absolute phase is proportional to the range difference between the primary and secondary acquisition from which the interferogram was formed. A similar view is adopted in optical interferometry \cite{creath85}. While the absolute interferometric phase is not observed directly, many authors assert that it needs to be known for estimating topography or displacements \cite{lombardini98,zebker98,rosen00}. But a general definition of the absolute phase is lacking. What is the absolute phase over a time period in spring in which snow melts and leaves flush, while coherence is partially retained? To answer this and similar questions, Sec. \ref{sec:wrappedunwrapped}--\ref{sec:definition} establish an observational definition of the absolute phase in terms of measurable quantities. This definition is applicable to any InSAR (Interferometric Synthetic Aperture) mode and does not require any knowledge about how the electromagnetic wave interacts with the observed target.

The observational definition of the absolute phase $\phia(\wb_2; \wb_1)$ postulates a continuous series of interferometric phase measurements that link the primary $\wb_1$ with the secondary $\wb_2$ configuration. Here, a configuration $\wb$ specifies the observation parameters such as the antenna position and time. Throughout the series of phase measurements, the primary configuration $\wb_1$ is held fixed, while the (intermediate) secondary configuration $\wb$ progresses from $\wb_1$ to the actual secondary configuration $\wb_2$ along a pre-specified absolute path. Unwrapping the interferometric phase as the secondary configuration is varied \cite{ghiglia98} produces the absolute phase. The absolute phase is congruent with the wrapped phase, differing only by the integer multiples of 2$\pi$ added during unwrapping. For differential InSAR with a temporal baseline, the absolute phase is determined by unwrapping the interferometric phase over time. The observational and the conventional definition agree for an ideal point target. 

The absolute phase thus defined introduces practical challenges for interpreting InSAR data. Specifically: 
\begin{enumerate}
\item It usually is not directly observable, as it requires a continuous series of observations. 
\item It is undefined if a phase singularity with vanishing coherence magnitude occurs at some point along the (unobserved) absolute path. I will refer to these situations as \textit{absolute phase singularities}.
\item It depends on the coherence along the entire path, not just the endpoints.
\end{enumerate}

\begin{figure}
		\centering
			\includegraphics{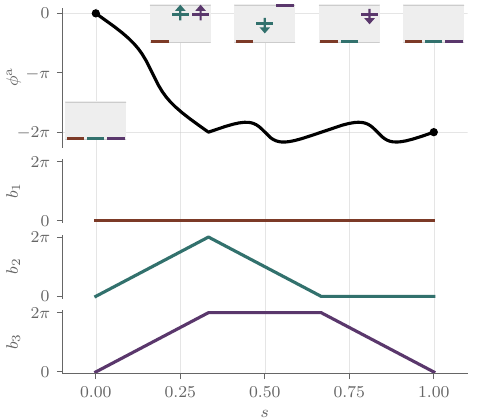}
		\caption{The differential InSAR absolute phase $\phia(s)$ as a function of dimensionless time $s$ for three surface segments. The InSAR observations do not resolve the segments and are continuous in $s$. The segments' normalized positions $b_1$, $b_2$, and $b_3$ are shown in the three bottom panels and also indicated in the insets. Model details can be found in Sec. \ref{sec:threeseg}.\label{fig:threesurfts}}
\end{figure}

An example of path dependence is shown in Fig. \ref{fig:threesurfts}, depicting the absolute phase $\phia(s)$ modeled for three moving surface segments as a function of dimensionless time $s$ for a fixed primary at $s = 0$. The surface segments, not resolved in the observations, are inspired by permafrost landforms comprised of multiple parts that undergo wavelength-scale differential movements \cite{zwieback24}. Segment 1 (red) stays stationary throughout, at position $b_1(s) = 0$. Conversely, segments 2 (teal) and 3 (blue) move up simultaneously for $s < 1/3$. Upward movement corresponds to a decreasing absolute phase, obtained by unwrapping over secondary time $s$. At secondary time $s = 1$, all segments have returned to their initial position, but the absolute phase $\phia(1) = -2\pi$. This nonzero value cannot be interpreted as a displacement or range difference, as the positions are identical at $s = 0$ and $s = 1$. Had the segments remained stationary, the absolute phase $\phia(s) = 0$ for all $s$. 

The loss of correspondence between absolute phase and range difference may seem undesirable, but it is not a quirk of the observational definition of the absolute phase. In differential interferometry the observationally defined absolute phase is the unique real-valued quantity that depends continuously on secondary time and is congruent with the wrapped phase (Sec. \ref{sec:differential}). In other words, proportionality to range difference is incompatible with a real-valued phase when there is path dependence. Such path dependence and related characteristics will be studied in examples from differential InSAR and across-track InSAR modes in Sec. \ref{sec:differential} and \ref{sec:across}.

The path dependence exhibited by complex targets controls how the absolute phase behaves across absolute phase singularities. In Sec. \ref{sec:patch}, it will be shown that the absolute phase can be patched continuously across absolute phase singularities if and only if there is no path dependence, provided certain continuity conditions hold. Using ideas from algebraic topology \cite{fulton97}, we will see how path dependence arises and, ultimately, how it complicates absolute phase reconstruction from actual observations.

Under what conditions can the absolute phase be recovered without continuous observations along the absolute path? A two-step approach --- phase unwrapping of multidimensional interferograms and integer offset determination --- is revisited in Sec. \ref{sec:reconstruct} in the context of absolute phase singularities. Unwrapping of, say, a spatial interferogram in two dimensions strives to reconstruct $\phia$ up to a constant offset that is an integer multiple of $2 \pi$. Absolute phase singularities present a challenge, as they are not directly observable but unwrapping across them can introduce reconstruction errors. To meet the challenge, the analyses establish conditions that guarantee correct reconstruction.

\section{Wrapped and unwrapped phase}
\label{sec:wrappedunwrapped}
\subsection{Wrapped phase}
\label{sec:wrapped}
\subsubsection{Definition}
The wrapped interferometric phase $\varphi$ is derived from two individual complex signals $u_1$ and $u_2$ by cross-multiplication \cite{bamler98}, viz.
\begin{align}
v &= \langle u_1  u_2^{\star} \rangle\,\\
  &= |v| \exp\left( \ju \varphi \right)\,.\label{sec:v}
\end{align}
Explicitly, $\varphi$ is the argument of $v$,
\begin{align}
\varphi = \arg(v).
\end{align}
The angular brackets $\langle \cdot \rangle$ denote an appropriate averaging operation, including an ensemble average (for modeling), a spatial multilooking operation, or no averaging (for single-look interferometry). I adopt the convention that $u_1$ -- the linear term -- is the primary signal and $u_2$ -- the conjugate-linear term involving complex conjugation $^{\star}$ -- is the secondary signal. 

The complex coherence $\gamma$ is defined as
\begin{align}
\gamma &= \frac{\langle u_1 u_2^{\star} \rangle}{\sqrt{\langle |u_1|^2 \rangle \langle |u_2|^2 \rangle}}\,.
\end{align}
The polar decomposition of this complex number 
\begin{align}
\gamma \equiv \rho \exp\left( \ju \varphi \right) \label{eq:gamma}
\end{align}
determines the wrapped phase $\varphi$ and the coherence magnitude $\rho \in [0, 1]$, a measure of the similarity between $u_1$ and $u_2$.

The wrapped phase $\varphi = \arg(\gamma)$ is only defined for $\rho > 0$, while a phase singularity occurs for $\rho = 0$. The engineering sign convention adopted throughout requires a positive sign in the complex exponential in \eqref{eq:gamma}. To convert to the physics sign convention with a negative sign, $\ju$ needs to be replaced by $-\iu$. An idealized measurement (e.g., no oscillator noise) produces error-free phase observations, as will be assumed in the theoretical analyses. This implies $\varphi = 0$ for identical primary and secondary signals. There is thus no gauge freedom in the interferometric phase \cite{dennis01}.

\subsubsection{Mathematical representation}
The wrapped phase is commonly represented by a number in an interval, but this introduces an arbitrary discontinuity in the complex plane. For the canonical representation in the half-open interval $(-\pi, \pi]$, the wrapped phase jumps by $\pm 2 \pi$ across the negative real axis \cite{dennis09}. Conversely, if we chose a different interval such as $(-2 \pi, 0]$, the discontinuity on the negative real axis would disappear, to be replaced by a different discontinuity or, in complex analysis language, a different branch cut of the complex logarithm. 

The wrapped phase $\varphi$ is naturally thought of as a point on the unit circle $S^1$ in the complex plane \cite{fulton97}. A $\varphi \in S^1$ is obtained by radially projecting a non-zero complex number $\gamma \in \Cn$ onto $S^1$, through the argument mapping $\arg: \mathbb{\Cn} \rightarrow S^1$. The circular nature arises from the periodicity of the complex exponential, as $\exp \ju \tilde{\varphi} = \exp \ju\left(\tilde{\varphi} + 2\pi N\right)$ for any $N \in \mathbb{Z}$ and any $\tilde{\varphi} \in \mathbb{R}$. 

An alternative representation capturing the circular nature uses equivalence classes. A $\tilde{\varphi} \in \mathbb{R}$ represents an equivalence class $[\tilde{\varphi}]$ of elements $\tilde{\varphi}' \in \mathbb{R}$ with $\exp \ju \tilde{\varphi} = \exp \ju \tilde{\varphi}'$, given by
\begin{align}
[\tilde{\varphi}] = \{\tilde{\varphi} + 2 \pi N \,\big|\, N \in \mathbb{Z}\}\,.
\end{align}
In abstract terms, such an equivalence class is an element of the quotient $\mathbb{R} / (2 \pi \mathbb{Z})$. 

\subsubsection{Dependence on variables}
The coherence, and in turn $\varphi$, commonly depend on variables. For defining the absolute phase, these variables describe InSAR configuration parameters such as the observation time and antenna position of the secondary acquisition, while the primary configuration is fixed. The configuration parameters are encoded by $\wb$ in a space $W$, assumed to be a subset of $\mathbb{R}^n$. For modeling, it can be convenient to fix the configuration parameters and introduce model parameters $\bb$ in a space $B$; they may for instance describe a scatterer's position. For describing observations such as a planar interferogram, we use $\xb \in X$ to denote spatial variables; for a two-dimensional interferogram, the space $X$ is $\mathbb{R}^2$. For now, we will focus on a configuration parameter space $W$.

We can think of a coherence field $\gamma(\wb)$ obtained through a complex-valued function $\gamma: W \rightarrow \mathbb{C}$ mapping $\wb \in W$ to a coherence $\gamma(\wb)$. The function $\gamma$ will be assumed to be continuous and smooth ($C^{\infty}$), apart from the discussion of discretely sampled data in Sec. \ref{sec:discrete}.

\begin{figure*}
		\centering
			\includegraphics{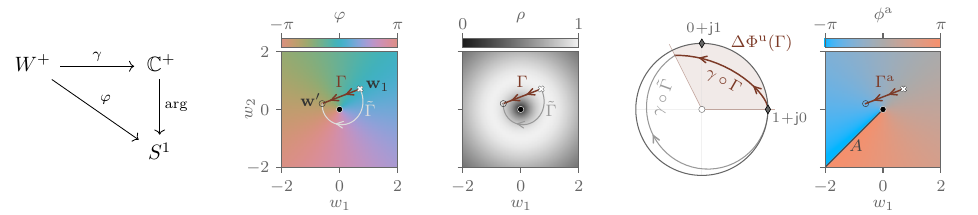}
		\caption{a) A coherence field $\gamma$ on $\Wn$ induces a wrapped phase field $\varphi$ through composition with the argument function. b--e) An example coherence field with b) wrapped phase and c) coherence magnitude, along with two paths $\Gamma$ and $\tilde{\Gamma}$ from $w_1$ to $w'$. d) The two paths are mapped into the complex plane through $\gamma$. e) The absolute phase, whose value at $w'$ is the unwrapped phase difference along $\Gamma$. \label{fig:walkthrough}}
\end{figure*}

The wrapped phase depends on $\wb$ as $\varphi(\wb) = \arg(\gamma(\wb))$. This expression does not make sense for phase singularities. We thus define the domain of $\varphi(\wb)$ as
\begin{align}
\Wn = W \setminus {\gamma^{-1}\left[\{0\}\right]}\,,
\end{align}
where the set difference $\setminus$ removes the pre-image of 0, denoted by $\gamma^{-1}\left[\{0\}\right]$. A point $\wb \in \Wn$ is in the pre-image of zero precisely if its image $\gamma(\wb) = 0$, or in other words, precisely if it is a phase singularity. In terms of functions, we have a mapping $\varphi: \Wn \rightarrow  S^1$ given by the composition $\varphi = \arg \circ \gamma$. Explicitly, $\varphi(\wb) = (\arg \circ \gamma)(\wb) = \arg\left(\gamma\left(\wb\right)\right)$, which is depicted graphically by the commutative diagram in Fig. \ref{fig:walkthrough}a). (The function $\gamma$ is restricted to $\Wn$.) The wrapped phase $\varphi$ inherits the $C^{\infty}$ smoothness from $\gamma$. 

An example of a smooth phase $\varphi$ derived from a smooth coherence $\gamma$ is shown in Fig. \ref{fig:walkthrough}b--c). While $\gamma$ is defined on $W \cong \mathbb{R}^2$ with coordinates $w_1$ and $w_2$, the phase is defined on $\Wn$, which excludes the origin. At the origin, the coherence magnitude $\rho = |\gamma| = 0$. 

\subsubsection{Phase singularities}
Phase singularities as zeros of $\mathbb{C}$-valued functions are ubiquitous in physics. They commonly occur as zeros of coherence functions (e.g., InSAR, optics) and as zeros of complex scalar wavefields (e.g., optics, acoustics, quantum mechanical wavefunctions). The latter are also called optical vortices or wavefront dislocations and feature in such phenomena as the Goos-H{\"a}nchen shift, seiches on a rotating Earth, or the ocean tides \cite{nye74,mortimer76,berry23}. For an overview of their history and properties, see \cite{berry09,dennis09,berry23}. Their topological characteristics, for instance how they form and annihilate in bifurcations when the complex field is disturbed, are discussed in \cite{freund95,berry06,adachi07}.

Phase singularities generically occur as points in the plane ($\mathbb{R}^2$). That is, almost all phase singularities are points \cite{berry09}, as that in Fig. \ref{fig:walkthrough}. Non-point singularities may arise from symmetry, such as the circles of vanishing amplitude in a circular aperture's Fraunhofer diffraction pattern. In $\mathbb{R}^3$, generic phase singularities form curves, including lines and loops \cite{huntley01,berry06}. In a general space $W$, their dimension is two less than that of $W$. Their codimension is two \cite{dennis09}, as both the real and imaginary part need to vanish.  
% for instance, the Fraunhofer diffraction pattern of a double slit or a circular aperture on a planar screen has lines of phase singularities, which a symmetry-breaking perturbation can splinter into isolated points \cite{dennis01}

A key characteristic of a phase singularity is that the phase can spiral around it \cite{berry23}, as in Fig. \ref{fig:walkthrough}b. Along a loop around a phase singularity, the phase changes by an integer multiple of $2 \pi$, this change being an unwrapped phase difference. 

\subsection{Unwrapped phase difference}
\subsubsection{Definition}
\label{sec:unwphasedef}
An unwrapped phase difference $\Delta\phi \in \mathbb{R}$ is an unambiguous measure of the difference between wrapped phases at two points connected by a path in a space $W$. To obtain that difference between a wrapped $\varphi(\wb) \in S^1$ at a point $\wb \in \Wn$ and another wrapped phase $\varphi(\wb') \in S^1$ at $\wb' \in \Wn$, unwrapping \cite{ghiglia98} integrates phase increments along a path connecting $\wb$ with $\wb'$. The result is not subject to the $2 \pi N$ ambiguity of the wrapped phase. Intuitively, this is because two sufficiently close points $\wb$ and $\wb + \mathrm{d}\wb$ will have almost the same phase due to the continuity of $\varphi(\wb)$, effectively removing the ambiguity \cite{fulton97}. The integration accumulates these unambiguous increments. 

For this unwrapping process, a connecting path $\Gamma$ from $\wb'$ to $\wb$ has to be specified. The rectifiable path $\Gamma$ is given by a continuous, piecewise smooth function from the unit interval $I \equiv [0, 1]$ to $\Wn$ such that $\Gamma(0) = \wb'$ and $\Gamma(1) = \wb$. 

Mathematically, the unwrapped phase difference $\Delta\phi(\Gamma)$ is given by \cite{ghiglia98}
\begin{align}
\Delta\phi(\Gamma) = \int_{\Gamma} \dphiW\,,\label{eq:phiux}
\end{align}
where $\dphiW$ is the differential one-form that measures phase changes. As a one-form, it takes an infinitesimal step $\mathrm{d}\wb$ as input and returns the infinitesimal phase change corresponding to $\mathrm{d}\wb$. For our purposes, it is sufficient to think of a differential $k$-form as an integrand in a $k$-dimensional integral \cite{weintraub14}. The one-form $\dphiW$ is integrated along a (one-dimensional) path, such as $\Gamma$ or $\tilde{\Gamma}$ shown in Fig. \ref{fig:walkthrough}b.

The one-form $\dphiW$ is often expressed as $\nabla \phi \cdot \mathrm{d}\wb$, with $\nabla \phi$ referred to as a gradient. However, in general \eqref{eq:phiux} depends on the path $\Gamma$, while it would not if $\nabla \phi$ were a true gradient. In the language of differential forms, $\dphiW$ is, in general, not exact \cite{weintraub14}. 

\subsubsection{Evaluation}
To evaluate the integral, we map the path $\Gamma$ into the punctured complex plane $\Cn = C \setminus {0}$. This is done through the coherence mapping $\gamma$, or alternatively through $v$. Representing the coherence as $\gamma = \xi + \ju \eta$, the change in phase relates to changes in $\xi$ and $\eta$ as
\begin{align}
\dphiC &= \frac{-\eta}{\xi^2 + \eta^2} \,\mathrm{d}\xi + \frac{\xi}{\xi^2 + \eta^2} \,\mathrm{d}\eta\,.
\end{align}
Intuitively, the coherence change $\mathrm{d}\gamma$ is projected onto the unit circle $S^1$. In the language of exterior calculus, $\dphi_{\Cn}$ is a differential one-form in the punctured complex plane. An equivalent expression is given by
\begin{align}
\dphi_{\Cn} = \frac{\mathrm{Im}\left(\gamma^{\star}\mathrm{d}\gamma\right)}{|\gamma|^2}\,.
\end{align}

Explicitly, the change of variables transforms \eqref{eq:phiux} to
\begin{align}
\Delta \phi(\Gamma) &= 
\int_{\gamma \circ \Gamma} \dphi_{\Cn} \label{eq:phiugamma} \\
&= \int_{0}^1 \left(\frac{-\eta(\Gamma(s))}{\rho(s)^2} \frac{\mathrm{d}\xi(\Gamma(s))}{\mathrm{d}s} + 
\frac{\xi(\Gamma(s))}{\rho(s)^2} \frac{\mathrm{d}\eta(\Gamma(s))}{\mathrm{d}s}\right)\,\mathrm{d}s\,, \nonumber
\end{align}

The evaluation is illustrated in Fig. \ref{fig:walkthrough}d) for the path $\Gamma$ from $\wb_1$ to $\wb'$, shown in red. Mapping this path into $\Cn$, the path $\gamma \circ \Gamma$ is obtained. The unwrapped phase $\Delta \phi(\Gamma)$ is the angle spanned by $\gamma \circ \Gamma$ as seen from the origin \cite{fulton97}. 

In the language of exterior calculus \cite{weintraub14}, the mapping $\gamma$ determines $\dphi_{\Wn} = \gamma^{\star}(\dphi_{\Cn})$ as the pullback from $\dphi_{\Cn}$. The pullback allows us to transfer evaluation of an infinitesimal phase change in the complex plane (through the one-form $\dphi_{\Cn}$) to $\Wn$ (through the one-form $\dphi_{\Wn}$).

\subsubsection{Path dependence of phase unwrapping}
\label{sec:pathunw}
For fixed endpoints, the unwrapped phase difference $\Delta\phi(\Gamma)$ can depend on the path $\Gamma$. In Fig. \ref{fig:walkthrough}b), paths $\Gamma$ and $\tilde{\Gamma}$ both start at $w_1$ and end at $w'$. In the complex plane (Fig. \ref{fig:walkthrough}d), $\gamma \circ \Gamma$ and $\gamma \circ \tilde{\Gamma}$ also have identical endpoints. However, the unwrapped phases $\Delta\phi(\Gamma)$ and $\Delta\phi(\tilde{\Gamma})$ are not equal, indicating path dependence. The difference of $2 \pi$ is due to the paths passing the origin on opposite sides. The path dependence ultimately derives from the phase singularity in $W$. If we were to first follow $\Gamma$ and then $\tilde{\Gamma}$ in the opposite direction (Fig. \ref{fig:walkthrough}b), we would encircle the phase singularity in a loop, leading to a total unwrapped phase around the loop of $-2 \pi$.

Any loop yields an unwrapped phase difference of $2 \pi N$, where $N \in \mathbb{N}$ \cite{fulton97}. The winding number $N$ is an integer that describes the number of phase cycles along the loop, or equivalently, how many times the loop mapped into the complex plane wraps around the origin. Alternative names for $N$ include residue (in InSAR), topological charge or topological quantum number (in physics). %In $\mathbb{R}^2$, a generic phase singularity such as that in Fig. \ref{fig:walkthrough} spirals once in either the counter-clockwise or clockwise direction around the phase singularity, corresponding to $N = 1$ and $N = -1$, but the sign convention varies \cite{dennis09,rosen00}. In InSAR, these are often referred to as positive or negative residues \cite{goldstein88, bamler98}. In physics, $N$ is commonly referred to as an optical charge or  \cite{berry09,berry23}. In higher dimensions, it is not straightforward to attribute a sign to a phase singularity because that would require a canonical loop orientation, but the phase unwrapping integral around a loop remains quantized.

\section{Defining the absolute phase}
\label{sec:definition}
\subsection{General definition}
I now propose a general definition of the absolute phase $\phi^{\mathrm{a}}(\wb_2)$ at $\wb_2 \in W$. It is an unwrapped phase difference in the configuration space $W$. The primary configuration $\wb_1 \in W$ is held fixed when evaluating the coherence and is also the starting point of the unwrapping integral. When the primary configuration $\wb_1$ is not clear from the context, we will write $\phi^{\mathrm{a}}(\wb_2; \wb_1)$. 

%The precise meaning of the space $W$ depends on the InSAR mode. In an observational setting, $\wb \in W$ describes the variables in the observational configuration that differ between the primary and secondary acquisition. In an ideal differential InSAR scenario, that will only be time. More generally, it can also include the antenna location or orientation. %For modeling the signals, $W$ will also contain parameters such as scatterer positions or total electron content of the ionosphere.

The absolute phase is unwrapped along a pre-specified absolute path $\Gamma^{\mathrm{a}}_{\wb_2} \in W$ that ends at $\wb_2$, representing the configuration at the secondary acquisition. The natural path $\Gamma_a$ handles path dependence of phase unwrapping by restricting attention to just one path. Its specification requires a choice and will depend on context. We will assume a reasonable choice has been made, deferring an in-depth discussion to later. %Let us assume We will assume $\Gamma_^{\mathrm{a}}_{\wb_2}$ exists and that it depends continuously on $\wb_2$, deferring discussion of its definition to the subsequent sections.

The defining equation of the absolute phase is
\begin{align}
\phia(\wb_2) \equiv \Delta\phi(\Gamma^{\mathrm{a}}_{\wb_2}) = \int_{\Gamma^{\mathrm{a}}_{\wb_2}} \dphi_{\Wn}\,.\label{eq:phia}
\end{align}

To illustrate, we return to Fig. \ref{fig:walkthrough} and define the absolute path $\Gamma^{\mathrm{a}}_{\wb_2}$ to be the straight-line path from $\wb_1$ to $\wb_2$. Thus, for $\wb_2 = \wb'$, the path $\Gamma$ shown in the figure is the absolute path, and $\phia(\wb') = \Delta \phi(\Gamma)$. 

\subsection{Absolute phase singularities}
I define an \textit{absolute phase singularity} to be any point $\wb_2$ for which the coherence vanishes somewhere along the path $\Gamma^{\mathrm{a}}_{\wb_2}$. The set of all such points, $A$, contains the phase singularities (for which the wrapped phase is undefined) as well as points for which the wrapped phase is defined. The absolute phase is not defined in $A$, but it is for all $\wb_2$ in
\begin{align}
\Wa = W \setminus A\,,
\end{align} 
provided $\Wa$ is path connected. Within $\Wa$, the absolute phase is a smooth function, provided $\Gamma^{\mathrm{a}}$ varies smoothly with $\wb_2$ and $\gamma(\wb_2)$ is a smooth function.

In Fig. \ref{fig:walkthrough}e, the absolute phase singularities $A$ lie on a ray extending from the phase singularity away from $\wb1$. For any $\wb \in A$, the absolute paths, namely the straight line paths from $\wb_1$, intersect the phase singularity. Across this phase singularity, the absolute phase jumps by $2 \pi$. This absolute phase singularity is thus similar to a branch cut of the complex logarithm. 

\subsection{Interpretation} 
The absolute phase $\phia(\wb_1, \wb_1) = 0$ for identical primary and secondary configuration parameters, by definition, assuming that the absolute path $\Gamma^{\mathrm{a}}_{\wb_1}$ is a constant path. % Need to explicitly state assumptions on absolute path

% Include sentence on Leibniz rule?

The absolute phase $\phia(\wb_2, \wb_1)$ is congruent \cite{fulton97,ghiglia98} with the wrapped phase $\varphi(\wb_2, \wb_1)$ in that 
\begin{align}
\arg\exp(\ju \phia(\wb_2, \wb_1)) = \varphi(\wb_2, \wb_1)\,. \label{eq:compatibility}
\end{align}
The two are equivalent modulo $2 \pi$, i.e., $\varphi(\wb_2, \wb_1) = [\phia(\wb_2, \wb_1)]$. While the wrapped phase changes sign when the primary and secondary acquisition are interchanged, $\varphi(\wb_2, \wb_1) = -\varphi(\wb_1, \wb_2)$, there is no such guarantee for the absolute phase. Assuming both $\phia(\wb_2, \wb_1)$ and $\phia(\wb_1, \wb_2)$ are defined, \eqref{eq:compatibility} implies that $[\phia(\wb_2, \wb_1)] = [-\phia(\wb_1, \wb_2)]$, but the integer multiple of $2 \pi$ will depend on the coherence mapping and the path $\Gamma^{\mathrm{a}}$. 

The definition of the path $\Gamma^{\mathrm{a}}: I \rightarrow W$ can be tailored to the mode. In this document, the convex combination,
\begin{align}
\Gamma^{\mathrm{a}}_{\wb_2}(s) = (1 - s) \wb_1 + s \wb_2\,, \label{eq:Gammaa}
\end{align}
is the default. It connects $\wb_1$ and $\wb_2$ through a straight line. It meets all three desiderata for $\Gamma^{\mathrm{a}}$: i) constancy of $\Gamma^{\mathrm{a}}_{\wb_1}$; ii) independence of $\gamma(\wb)$, and iii) smooth dependence on the endpoint $x_2$. Its definition makes sense for star-shaped $W$ (with center $\wb_1$). For more general $W$, different choices of $\Gamma^{\mathrm{a}}$ are required. If the absolute path has discontinuities, these are then included in the absolute phase singularities $A$, implying that the absolute phase is not defined on such points and rescuing continuity of the absolute phase in $W \setminus A$.

\subsection{Analogy}
A familiar analogy unrelated to interferometry illustrates these concepts: The offset in the time of day on a spherical Earth relative to a standard time. The time zones are a discretized variant of this offset \cite{dennis01}. The wrapped phase for any point on the Earth's surface corresponds to the difference between local solar time and a reference time, with a local solar time of zero when the sun is lowest below the horizon. (For simplicity, I will neglect the tilt of the Earth's axis and the eccentricity of its orbit, thus equating mean and apparent solar time.) The sine of the amplitude of the sun's elevation angle relative to the local horizon may be associated with a coherence magnitude, with the Earth's poles as phase singularities \cite{dennis01}. 

An absolute phase is induced by definition of an absolute path with an initial point of 0$^{\circ}$ latitude and 0$^{\circ}$ longitude. Define the absolute path to be the shortest among all paths that first run along the equator and then along a line of longitude, choosing to go East first for points on the 180$^{\circ}$ meridian. The absolute phase in this case is restricted to the interval $(-\pi, \pi]$ and establishes the notion of a calendar day for the entire Earth. The jump by 1 day in time (or $2 \pi$ in phase) identifies the 180$^{\circ}$ longitude line as a simplified international date line, or an absolute phase singularity based on the convention from the previous subsection. A different absolute path -- for instance, only going west along the equator -- would induce a different absolute phase or calendar day. More generally, the unwrapped phase along a path corresponds to the time difference between local solar time and the time shown by the clock of a traveler who has traversed the path and who initially synchronized their clock. A traveler going around the world in eighty days in an eastward loop perceives a time difference of one day \cite{verne73}. 

\section{Differential InSAR absolute phase}
\label{sec:differential}
\subsection{Definition}
\label{sec:differentialdefinition}
In an ideal case, only time distinguishes the secondary from the primary configuration. Keeping the primary acquisition at time $t_1$ fixed, we can let $W \cong \mathbb{R}$ and $\wb \equiv t$. From \eqref{eq:phia} and \eqref{eq:Gammaa}, the differential InSAR absolute phase at a secondary time $t_2$ is defined by
\begin{align}
\phi^{\mathrm{a}}(t_2) &= \int_{t_1}^{t_2} \dphi_{t}\\
&= \int_{t_1}^{t_2} 
\left(\frac{-\eta(t)}{\rho(t)^2} \frac{\mathrm{d}\xi(t)}{\mathrm{d}t} + 
\frac{\xi(t)}{\rho(t)^2} \frac{\mathrm{d}\eta(t)}{\mathrm{d}t}\right)\,\mathrm{d}t\,.\label{eq:phiadiff}
\end{align}
The coherence $\gamma(t) = \xi(t) + \ju \eta(t)$ is evaluated for a fixed primary at $t_1$ and variable intermediate secondary acquisition at $t$. It can be convenient to use a dimensionless time $s = (t - t_1) / (t_2 - t_1)$, in which case the limits are $0$ and $1$. Then, $\phi^{\mathrm{a}}(s)$ for $0\leq s < 1$ is the absolute phase for ``intermediate'' secondary acquisitions prior to the ``final'' secondary acquisition at $s = 1$.
%if real object has the same $\varphi(s)$ as a point scatterer, it will have the same absolute phase as that point scatterer.

In differential InSAR, the absolute path basically requires no choice. The observational definition of the absolute phase in \eqref{eq:phiadiff} is the unique definition of a phase-like quantity in $\mathbb{R}$ that fulfills three intuitive axioms, for fixed $t_1$ and continuous $\gamma(t_2)$. Axiom 1: it depends continuously on secondary time $t_2$; axiom 2: it is congruent with the wrapped phase $\varphi(t_2)$ for all $t_2$; and axiom 3: it is zero for $t_1 = t_2$. The statement then follows from Proposition 1.34 in \cite{hatcher01}; in the theory of covering spaces, it is an application of the unique lifting property of a mapping into $S^1$. Axioms 1 and 2 determine the phase-like quantity up to an integer multiple of $2 \pi$, while axiom 3 fixes this integer.

Measuring this absolute phase requires continuous observations. Conversely, most real-world data obtained at discrete measurement times do not provide direct observation of the absolute phase.

\subsection{Modeling framework}
\label{sec:diffmodeling}
To predict the absolute phase in a modeling scenario, consider model parameters $\bb$ in some space $B$. The parameters $\bb$ describe the system at all times, so that the coherence is modeled as $\gamma(\wb_2, \wb_1, \bb)$. The absolute phase $\phia(\xb_2, \xb_1, \bb)$ also depends on $\bb$, by evaluation of \eqref{eq:phia} for fixed $\bb$. 

For theoretical analyses, a simplified modeling framework that applies to differential InSAR will be used. The parameters $\bb(t)$ describe the system as a function of time $t$; for instance, the position and radar cross section of multiple scatterers. The coherence between times $t$ and $t'$ is modeled as $\gamma(\bb(t), \bb(t'))$ and assumed to be a smooth function. If we fix the primary and secondary observation times $t_1$ and $t_2$, we can use a path $\Gamma$ in $B$ to describe how the system evolves from $t_1$ to $t_2$. The simplified modeling framework restricts the evolution for a fixed $\bb_1$ and imposes that the system reach $\bb_2$ along a prespecified absolute path $\Gamma^{\mathrm{a}}_{\bb_2}$ in parameter space $B$. Within this modeling framework, the absolute phase in parameter space is
\begin{align}
\phia_{\Bn}(\bb_2) = \int_{\Gamma^{\mathrm{a}}_{\bb_2}} \dphi_{\Bn}\,,
\end{align}
where the differential one-form $\dphi_{\Bn}$ is analogous to $\dphi_{\Wn}$. This simplified framework is mathematically convenient because $B$ now has the same role as $W$. In particular, we call $\bb_2$ an absolute phase singularity if the phase is undefined somewhere along $\Gamma^{\mathrm{a}}_{\bb_2}$, at an intermediate time. The domain of $\phia_{\Bn}$ is denoted $\Ba$. As for $\phia$, we will adopt the convex combination absolute path \eqref{eq:Gammaa}. For instance, if $\bb$ encodes scatterer positions, the scatterers will move at a uniform rate from $\bb_1$ to $\bb_2$.

\subsection{Examples}
\subsubsection{Point scatterer}
For a point scatterer, the absolute phase defined through \eqref{eq:phiadiff} is proportional to the difference in optical path length, as expected \cite{bamler98}. Assuming a monostatic configuration, zero spatial baseline, the plane wave approximation and no atmospheric or other phase screens, the single-look coherence 
\begin{align}
\gamma_{\mathrm{point}}(t) = \exp\left(\ju 2 k \left[R(t) - R(t_1)\right]\right)\,, \label{eq:phipoint}
\end{align}
where $k$ is the scalar wavenumber, $R$ the range (line-of-sight distance to the scatterer), and the primary acquisition at $t_1$ is fixed. Multilook observations of a non-decorrelating, uniformly moving surface are also described by \eqref{eq:phipoint}. Previously \cite{bamler98,rosen00}, the absolute phase of such scatterers was taken as 
\begin{align}
\phi^{\mathrm{a}}_{\mathrm{point}}(t_2; t_1) = 2 k \left[R(t_2) - R(t_1)\right]\,. \label{eq:distance}
\end{align}
This expression agrees with \eqref{eq:phia}, as 
\begin{align}
\dphi_{t}(\mathrm{d}t) = 2 k \frac{\mathrm{d}R}{\mathrm{d}t} \,\mathrm{d}t = 2 k \,\mathrm{d}R\,
\end{align}
is an exact form, so the integral only depends on the range difference $R(t_2) - R(t_1)$ and not on the path $R(t)$.

\subsubsection{Three surface segments}
\label{sec:threeseg}
\begin{figure}
		\centering
			\includegraphics{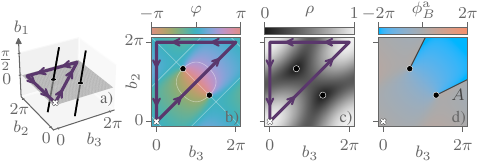}
		\caption{Differential InSAR model of three surface segments with positions $\bb = [b_1, b_2, b_3]$. a) The purple loop with basepoint $\bb_1 = [0, 0, 0]$ (white cross) in parameter space describes the temporal evolution from Fig. \ref{fig:threesurfts}, it encircles a phase singularity (black line) in the plane $b_1 = 0$. b--d) Wrapped phase with white contours indicating vanishing real and imaginary part (b), coherence (c) and absolute phase in the modeling framework of Sec. \ref{sec:diffmodeling} (d) shown in the plane $b_1 = 0$ for fixed primary $\bb_1$, along with phase singularities (black dots) and absolute phase singularities (gray lines).}\label{fig:threesurf}
\end{figure}

The three surface segments from the introduction differ from a single surface or point scatterer in that a cyclic change can induce a non-zero absolute phase. To model the system, let the state $\bb = [b_1, b_2, b_3]$ encode the position of the three segments, which are not resolved in the observations. Assuming the returns from different segments are uncorrelated and of equal strength \cite{zwieback24}, let $b_i(s)$ be the position of surface $i$ at dimensionless time $s$, nondimensionalized such that the modeled population coherence is 
\begin{align}
\gamma(s) = \frac{1}{3} \sum_{i=1}^3 \exp^{- \ju \left(b_i(s) - b_i(0)\right)}\,.
\end{align}
Each segment starts out at $b_i(0) = 0$ for the fixed primary acquisition at $s = 0$ and acts like a non-decorrelating point-like scatterer from \eqref{eq:phipoint}. Fig. \ref{fig:threesurf}a shows a loop that corresponds to the cyclic change in surface positions from Fig. \ref{fig:threesurfts}. As all segments return to their initial position at $s = 1$, a nonzero value of the absolute phase $\phia(1) = -2 \pi$ is obtained.

The nonzero absolute phase for the cyclic change demonstrates that the absolute phase is not proportional to a range difference, which may seem counterintuitive. In the case of the three surface segments, an obvious way to measure the range difference with respect to $\bb_1 = [0, 0, 0]$ is the average of $\bb = [b_1, b_2, b_3]$. A weighted average would be another option. In general, let us denote by $\Delta r(\bb)$ any measure of the range difference that vanishes whenever $\bb = \bb_1 \equiv [0, 0, 0]$ and that is continuous in $\bb$. To combat the counter-intuitive nature of the absolute phase defined through \eqref{eq:phiadiff}, couldn't we define an alternative absolute phase that is proportional to $\Delta r$? The answer is no, as long we require that absolute phase to be congruent with the wrapped phase (making it a phase-like quantity), to depend continuously on $\bb(s)$ (as $\Delta r$ does), and to equal zero when the surface positions do not change (as $\Delta r$ does). For any continuous temporal trajectory $\bb(s)$, these three requirements entail (Sec. \ref{sec:differentialdefinition}) the observational definition of the absolute phase in \eqref{eq:phiadiff}. The non-zero value $\phia(1) = -2 \pi$ for the cyclic change implies that the absolute phase cannot be proportional to any $\Delta r$. To develop insight into the origin of the loss of proportionality to a range difference, we next consider the phase in parameter space.

The nonzero absolute phase around the loop relates to the loop enclosing a phase singularity in parameter space, one of the two dark lines in Fig. \ref{fig:threesurf}a). In the plane $b_1 = 0$, the wrapped phase $\phi(\bb)$ for fixed primary $\bb_1 = 0$ in Fig. \ref{fig:threesurf}b spirals around the phase singularity, such that the unwrapped phase around the loop gives the absolute phase $\phia(1) =- 2 \pi$. As $\rho > 0$ along the loop (Fig. \ref{fig:threesurf}c), this absolute phase is well defined. 

The phase singularities induce absolute phase singularities in the simplified modeling framework of Sec. \ref{sec:diffmodeling}. While the loop does not fit into this framework, let us impose the framework by restricting trajectories to the line segment from $\bb_1 = 0$ to secondary $\bb$. The absolute phase in parameter space, $\phia_B$ is shown in Fig. \ref{fig:threesurf}d). The absolute phase singularities $A \in B$ in the plane $b_1 = 0$ are rays extending from each phase singularity, across which the absolute phase jumps by $2 \pi$. They leave no trace in the wrapped phase. In Sec. \ref{eq:patching}, it will be shown that the discontinuous jumps arise from the path-dependence of the phase one-form illustrated by the loop encircling the phase singularity. 

\subsubsection{Intermittent decorrelation}
Intermittent decorrelation, as may be caused by seasonal vegetation or snow, is present in the following model of a single-look signal $u(s)$ as a function of dimensionless time $s \in [0, 1]$:
\begin{align}
u(s) = u_{\mathrm{p}} \exp(-\ju \pi s) + f(s) \exp(\ju \pi s) u_{\mathrm{n}}\,. \label{eq:decorrelation}
\end{align}
The persistent contribution $u_{\mathrm{p}}$ undergoes an $s$-dependent phase change. The nuisance contribution $u_{\mathrm{n}}$ induces intermittent decorrelation: The modulation by $f(s) = 16 s (1 -s)$ is chosen to vanish at $s = 0$ and $s = 1$. Both $u_{\mathrm{p}}$ and $u_{\mathrm{n}}$ are drawn independently from a circularly-symmetric standard complex normal distribution and remain perfectly coherent \cite{bamler98}.

\begin{figure}
		\centering
			\includegraphics{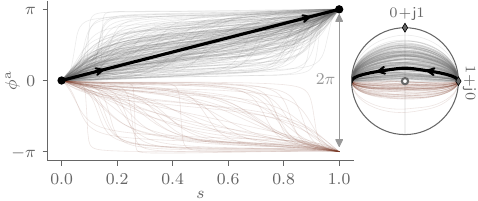}
		\caption{Modeled differential InSAR absolute phase for intermittent decorrelation from \eqref{eq:decorrelation} with $\beta = 16$. The thick black line represents the ensemble, the thin lines the speckle-affected realizations, color coded depending on $\phia(1)$. The polar plot on the right shows the coherence $\gamma$, the outer circle corresponding to $\rho = 1$.} \label{fig:decorrelation}
\end{figure}

The absolute phase $\phi^{\mathrm{a}}(s)$ obtained from the ensemble coherence for fixed primary acquisition at $s = 0$ is shown by a thick black line in Fig. \ref{fig:decorrelation}. Due to the persistent contribution, $\phi^{\mathrm{a}}$ increases linearly with $s$, reaching $\pi$ at $s = 1$. In the polar plot on the right, the black line starts at $\rho(0) = 1$, $\varphi(0) = 0$, moving to the left with increasing $s$ as the phase increases while the coherence magnitude $\rho$ first drops before recovering. For $s = 1$, the wrapped phase $\varphi(1) = \pi$ would be congruent with an absolute phase of $-\pi$ or $\pi$, but $\phi^{\mathrm{a}}(1) = \pi$ due to the trajectory.

The impact of speckle on the absolute phase is illustrated by the thin lines in Fig. \ref{fig:decorrelation}, corresponding to multiple realizations of 8-look observations. Most realizations -- shown in gray -- end at an absolute phase $\phi^{\mathrm{a}}(1) = \pi$, while some -- shown in red -- terminate at $-\pi$. In the polar plot, the red lines are those that pass the imaginary axis below the origin. Despite the perfect coherence $\rho = 1$ and error-free $\varphi = \pi$ observations for $s = 1$, the absolute phase differs between realizations due to the intermittent decorrelation. The absolute phase is not amenable to ensemble averaging. If the realizations were from different locations within a homogeneous region, the absolute phase would be discontinuous.

\subsubsection{Tree canopy observations}
\label{sec:borealscat}
\begin{figure*}
		\centering
			\includegraphics{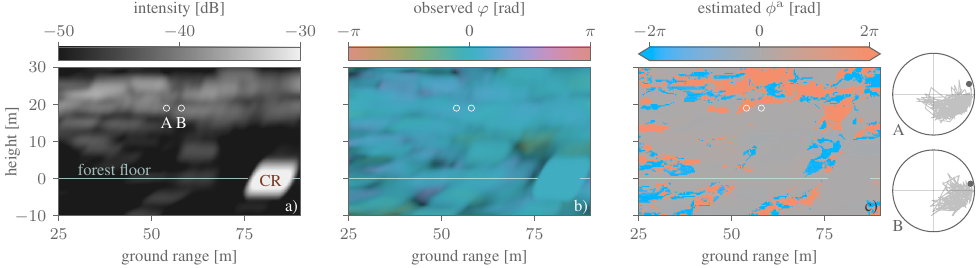}
		\caption{BorealScat tomographic observations of a forest canopy. a) The multilooked intensity on 2018-07-24 00:00 decreases downward from canopy top at $\sim$25 m, except the subcanopy corner reflector (CR). b) 24-hour interferogram (July 24 to July 25, midnight) with wrapped phase (hue) at coherence magnitude (brightness). c) Absolute phase estimate from successive 5-min interferograms, with points A and B shown as markers in a--c) differing by $2 \pi$. The path in $\mathbb{C}$ traced by A and B over the 24-hour interval is shown in the insets. \label{fig:borealscat}}
\end{figure*}

Transient coherence drops are also observed in the L-band BorealScat tomographic time series of a forest canopy in Remningstorp, southern Sweden \cite{monteith22,esa23}. Fig. \ref{fig:borealscat}a) shows an uncalibrated tomogram from July 24, 2018, derived from the tower-mounted antenna array \cite{ulander18}, with the strongest return from the top of the canopy 20--25 m above the forest floor. The upper canopy has $\rho \gtrsim 0.8$ in the unreferenced one-day interferogram in b) and the wrapped phase $\varphi$ is largely within $\pi/8$ of zero.

The estimated absolute phase in Fig. \ref{fig:borealscat}c) shows $2 \pi$ discontinuities, despite the near uniformity of the 24-hour interferogram. The absolute phase was estimated by approximating the integral \eqref{eq:phiadiff} by adding the wrapped phase differences over successive 5-minute intervals, the sampling interval of the BorealScat observations. Upper-canopy points A and B differ by $2 \pi$ in this estimated absolute phase, although the 24-hour complex coherence is similar. The complex plane insets shows that the coherence timeseries (for fixed primary) differ between these points, with only that of A encircling the origin during a period of low coherence. The low-coherence period occurred during the day, characterized by windier conditions and lower plant water content. 

The interpretation of the estimated absolute phase is challenging. The first challenge is unwrapping errors along the absolute path, despite the short 5-min sampling interval. Second, the absolute phase of a speckle-affected realization differs from that of the hypothetical speckle-free underlying process (Fig. \ref{fig:decorrelation}). Under the assumption that the observations sample a coherence that varies smoothly in space and time and that the phase is adequately sampled in time, the absolute phase jumps of $2 \pi$ in Fig. \ref{fig:borealscat}c) coincide with absolute phase singularities for which the coherence vanished at an earlier, intermediate time. The jumps reflect the sensitivity of phase unwrapping to perturbations at low coherence, and their occurrence in integer multiples of $2 \pi$ is due to the congruence of the absolute phase with the (near-zero) wrapped phase. Both characteristics impede interpreting the observed absolute phase as a range difference.

\section{Across-track InSAR absolute phase}
\label{sec:across}
\subsection{Definition}
For monostatic InSAR with a non-zero across-track spatial baseline \cite{madsen93, rosen00}, we will distinguish single-pass and repeat-pass observations in the definition of $\Gamma^{\mathrm{a}}$ through \eqref{eq:Gammaa}. In the single-pass case with $t_1 = t_2$, the primary and secondary configuration only differ in the spatial baseline. For a fixed line-of-sight direction, consider a simplified two-dimensional scenario in which the parallel baseline is zero. Denoting the perpendicular antenna position by $b$, we have $\wb = b$ and the perpendicular baseline $b_2 - b_1$. The path $\Gamma^{\mathrm{a}}$ traces antenna positions from $b_1$ to $b_2$. In the repeat-pass case with $t_1 \neq t_2$, we have $\wb = [t, b]$. The natural path $\Gamma^{\mathrm{a}}$ from \eqref{eq:Gammaa} uniformly and simultaneously increases the temporal and spatial baseline. We can think of it as the antenna moving slowly from primary to secondary position. In both single-pass and repeat-pass cases, a frequency offset $\Delta f$ applied to mitigate surface decorrelation \cite{bamler98} could be incorporated by appending $\Delta f$ to $\wb$. Conversely, the following examples focus on single-look observations of point-like scatterers.

\subsection{Examples}
\subsubsection{Single-pass: Point scatterer}
The single-pass absolute phase of an ideal point scatterer equals the difference in propagation phases \cite{madsen93}. Denoting the distance between the scatterer and the antenna at $b$ by $R(b)$, the single-look-complex signal, normalized to unit magnitude, is $u_{\mathrm{point}}(b) = \exp\left(- \ju 2 k R(b)\right)$. According to \eqref{eq:Gammaa}, the absolute phase is obtained from the series of interferograms with baseline $s \left(b_2 - b_1\right)$, where $s$ is increased from 0 to 1. The integral yields
\begin{align}
\phi^{\mathrm{a}}_{\mathrm{point}} = - 2 k \left( R(b_1) - R(b_2)\right)\,.\label{eq:singlepasspoint}
\end{align}
This is just the difference in the propagation phases and is proportional to the range difference $R(b_1) - R(b_2)$. It coincides with prior work that stipulated the absolute phase to be equal to this difference \cite{madsen93,rosen00}.

\begin{figure}
		\centering
			\includegraphics{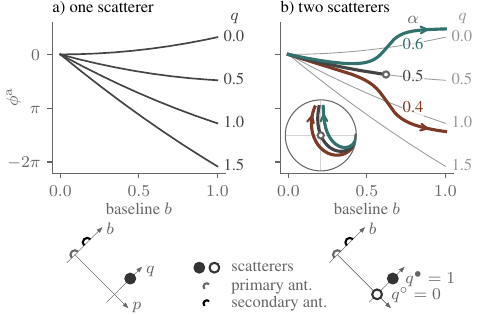}
		\caption{Single-pass absolute phase for single-look observations of a) one or b) two point scatterers versus across-track baseline $b$, non-dimensionalized such that the basline Fresnel number $F_b = 0.16$. In a), the lines correspond to across-track positions $q$ (nondimensionalized). In b), the two scatterers are located at $q^{\circ} = 0$ and $q^{\bullet} = 1$; the heavy lines correspond to different weights $\alpha$.} \label{fig:monostatic}
\end{figure}

The absolute phase of a point scatterer is essential for height determination \cite{madsen93}. For illustration purposes, we determine the view-perpendicular position $q$, which can be converted to a height, and keep the along-axis position $p$ fixed. To simplify notation, let $q$ be normalized (with $q = 1$ corresponding to a target Fresnel number $F_q = 1$), let $b_1 = 0$ and interpret $b_2 \equiv b$ as the spatial baseline, expressed in arbitrary units. Fig. \ref{fig:monostatic}a) shows $\phia$ as a function of $b$ for different values of $q$. For any baseline $b$, $\phia$ depends on $q$. Conversely, an estimate of $q$ is obtained by matching the observed $\phia$ to the point scatterer's modeled $\phi^{\mathrm{a}}_{\mathrm{point}}$. This estimate is commonly interpreted as a phase center position corresponding to the observed phase \cite{rosen00}.

\subsubsection{Single-pass: Two point scatterers}
\label{sec:twopoint}
Two point scatterers are the simplest example of extended targets, which include vegetation or ice. The single-look-complex signal is modeled as
\begin{align}
u(b) = \alpha u_{\mathrm{point}}^{\circ}(b) + (1 - \alpha) u_{\mathrm{point}}^{\bullet}(b)\,,
\end{align}
where $u_{\mathrm{point}}^{\circ}$ and $u_{\mathrm{point}}^{\bullet}$ are the contributions from point scatterers located at $q^{\circ}$ and $q^{\bullet}$, respectively. The weight parameter $\alpha$ determines the relative strength. This coherent model only accounts for the direct return from each scatterer. 

The two-scatterer absolute phase is plotted as a function of baseline $b$ with heavy lines in Fig. \ref{fig:monostatic}b). For equally strong scatterers ($\alpha = 0.5$; black), the absolute phase $\phi^{\mathrm{a}}(b)$ for small baselines $b \lesssim 0.6$ agrees with that of a point scatterer located at $q = 0.5$, halfway between the two scatterers. However, the inset plots shows decreasing magnitude with increasing $b$ (direction of arrows) arising from negative interference, while the magnitude of a single scatterer would be independent of baseline. For a singular baseline $b_{\mathrm{s}} \approx 0.6$, the circle in both plots demarcates where the return is numerically zero. At this singularity, $\phi^{\mathrm{a}}$ of a point scatterer at $q = 0 = q_{\circ}$ deviates from that of a point scatter at $q = 1 = q^{\bullet}$ by $\pi$. While $\varphi$ is well-defined for $b > b_{\mathrm{s}}$, neither the absolute phase nor the phase center position are. 

The absolute phase of pair of unequally bright scatterers is influenced by the singularity. For the case where the hollow scatterer at $q^{\circ} = 0$ is brighter ($\alpha = 0.6$; green), $\phi^{\mathrm{a}}(b)$ exceeds that of a point scatterer at $q = 0$ for $b \gtrsim b_{\mathrm{c}}$. The phase center for $b = b_{\mathrm{max}}$ is at $q < 0$. Conversely, that for $\alpha = 0.4$ (red) exceeds $q = 1$, the position of the filled scatterer. While the wrapped phases for $b = b_{\mathrm{max}}$ are seen to be similar in the polar plot, the absolute phases deviate by more than $\pi$. 

\subsubsection{Repeat-pass: Two point scatterers}
In the repeat-pass case, the two point scatterers can move and the scattering weight $\alpha$ can change between the primary $t_1$ and secondary time $t_2$. Using normalized time $t$ and baseline $b$, we have $t_1 = 0$ and $b_1 = 0$ for the primary and $t_2 = 1$ and $b_2 = 1$ for the secondary acquisition.

\begin{figure*}
		\centering
			\includegraphics{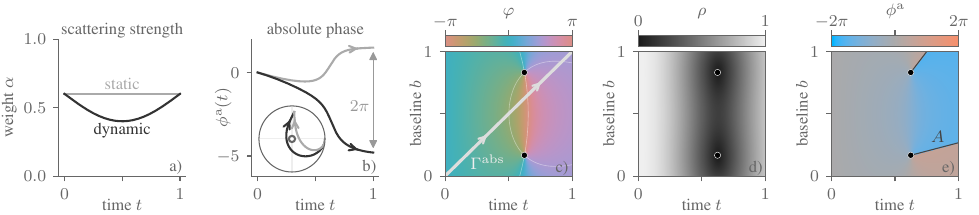}
		\caption{Repeat-pass InSAR model of two point scatterers with scattering weight $\alpha$. The weights $\alpha(t)$ for the dynamic and a static case (a) correspond to absolute phases (b) that differ by $2 \pi$ at $t = 1$, as their $\gamma$ pass the origin at opposite sides (inset). The wrapped phase $\varphi$ (c), coherence magnitude $\rho$ (d), and absolute phase $\phia$ (e) for variable secondary time $t$ and baseline $b$, and dynamic $\alpha(t)$.}\label{fig:dynamicalpha}
\end{figure*}

Fig. \ref{fig:dynamicalpha}a--b shows a dynamic scenario, in which the scattering strength $\alpha(t)$ depends on time $t$. Starting at $\alpha(0) = 0.6$ at the primary time $t_1 = 0$, it dips below 0.6, returning to 0.6 for secondary time $t_2 = 1$. The absolute phase $\phia(1)$ at the endpoint of the absolute path $\Gamma^{\mathrm{a}}$ to $\wb_2 = [1, 1]$ is approximately -5 in this dynamic case, while the static case with constant $\alpha = 0.6$ is greater by $2 \pi$. The difference in absolute phase stems from the observation in the polar plot inset that the coherence trajectories along the absolute path pass the origin on opposite sides. Conversely, the complex coherence and thus $\varphi$ at the $\Gamma^{\mathrm{a}}$ endpoint are identical for the dynamic and static case. Fig. \ref{fig:dynamicalpha}c--d) show $\varphi$ and $\rho$ for the dynamic case as a function of variable secondary time $t$ and baseline $b$. The absolute path, which is independent of the observed target, passes between two phase singularities, around which the phase $\varphi$ spirals by $2 \pi$. 

The absolute phase $\phia$ for variable secondary time $t$ and baseline $b$ is shown in Fig. \ref{fig:dynamicalpha}e, evaluated along the path \eqref{eq:Gammaa}. Absolute phase singularities $A$ radiate outward from either phase singularity. The absolute phase jumps by $2 \pi$ across these absolute phase singularities. 

\section{Patching absolute phase singularities}
\label{sec:patch}
Fig. \ref{fig:dynamicalpha}e features absolute phase singularities across which the absolute phase jumps by a non-zero integer multiple of $2 \pi$. These cannot be patched. By patching, I mean extending $\phia$ continuously from $\Wa$ to $\Wn$ to obtain a function $\psi: \Wn \rightarrow \mathbb{R}$ such that $\psi(\wb) =\phia(\wb) $ for all $\wb \in \Wa$ and $\mathrm{d}\psi = \dphi_{\Wn}$ in $\Wn$. Loosely speaking, the goal is to add an integer multiple of $2 \pi$ to the wrapped phase at absolute phase singularities such that the resulting continuous function $\psi$ matches $\phia$ off the absolute phase singularities. More precisely, we can only hope to patch those absolute phase singularities for which the wrapped phase is defined, i.e., $A' = A \setminus {\gamma^{-1}\left[\{0\}\right]}$. 

\begin{figure}
		\centering
			\includegraphics{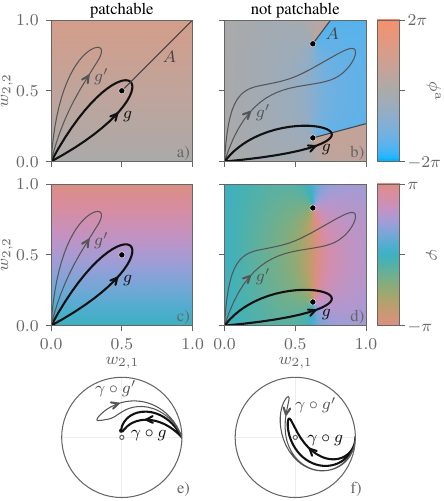}
		\caption{Two examples of absolute phase singularities for varying $\wb_2 = [w_{2, 1}, w_{2, 2}]$ with fixed $\wb_1 = [0, 0]$ for: a) patchable and b) unpatchable (same example as Fig. \ref{fig:dynamicalpha}). The corresponding wrapped phases $\varphi$ are shown in c) and d), respectively, and the loops projected into the complex plane in e) and f). \label{fig:patchability}}
\end{figure}

An example of a patchable absolute phase singularity is shown in Fig. \ref{fig:patchability}a. The absolute phase  $\phia$ is the same on either side of the absolute phase singularity $A$ for fixed $w_{2, 2}$, allowing us to extend $\phia$ continuously. Conversely, the absolute phase singularities in Fig. \ref{fig:patchability}b cannot be patched. 

When can we patch the absolute phase singularities? Patchability turns out to be closely connected to path independence of line integrals $\int \dphi_{\Wn}$. 

\subsection{Patching and path independence}
\label{eq:patching}
The absolute phase can be extended across absolute phase singularities $A'$ if and only if the integral $\int_{\Gamma} \dphi_{\Wn}$ does not depend on the path $\Gamma \in \Wn \subset \mathbb{R}^n$ from fixed $\wb_1$ to all $\wb_2$. It is assumed that $\gamma(\wb_2)$ is smooth and $\Wn \cong S \subset \mathbb{R}^n$ is path connected and locally path connected. The statement is closely related to the fact that a closed one-form $\omega$ on a subset of $\mathbb{R}^n$ is exact -- i.e., there exists a function $f$ such that $\omega = \mathrm{d}f$ -- if and only if the line integral $\int_{\Gamma} \omega$ is path independent (Theorem 6.4.12 in Chapter IV, \cite{weintraub14}). 

To see the \textit{if} part, define $\psi(\wb_2)$ as the integral of $\dphi_{\Wn}$ along any path in $\Wn$ from $\wb_1$ to $\wb_2$. The function $\psi$ is well defined due to path independence and continuous. For $\wb_2 \in \Wa$, $\psi(\wb_2) = \phia(\wb_2)$. However, $\psi$ can be evaluated at absolute phase singularities $A'$. Basically, for $\wb_2 \in A'$, we have replaced the problematic $\Gamma^{\mathrm{a}}$ by another path. It can be shown that $\dphi_{\Wn} = \mathrm{d}\psi$ in $\Wn$, as required \cite{weintraub14}.

To see the \textit{only if} part, note that the existence of a patched $\psi$ implies that $\dphi_{\Wn} = \mathrm{d}\psi$ is exact on $\Wn$, from which path independence follows. Exactness, together with $\psi(\xb_1) = 0$, further implies that the patched $\psi$ is unique.

Our next goal is to identify path dependence and understand how it can break down across absolute phase singularities.

\subsection{Loops for characterizing path dependence}
To identify path dependence of the phase integral, the concept of loops introduced in Sec. \ref{sec:pathunw} is convenient \cite{hatcher01}. For arbitrary $\wb_2$, we consider two paths $\Gamma$ and $\Gamma'$, i.e., functions $[0, 1] \rightarrow \Wn$ with $\Gamma(0) = \Gamma'(0) = \wb_1$ and $
\Gamma(1) = \Gamma'(1) = \wb_2$. To form a loop, consider the inverse path $\bar{\Gamma'}$, defined by $\bar{\Gamma'}(s) = \Gamma'(1 - s)$: the traversal is in the opposite direction, from $\wb_2$ to $\wb_1$. Concatenation $\Gamma \cdot \bar{\Gamma'}$ yields a loop that starts and ends at $\wb_1$, the loop's basepoint. This construction \cite{fulton97} is useful because the integral in \eqref{eq:phia} is path independent for all $\wb_2$ if and only if
\begin{align}
\int_{g} \dphi_{\Wn} = 0 \quad \text{for all loops $g$ in $\Wn$ with basepoint $\wb_1$}\,. \nonumber
\end{align}
If it holds, the absolute phase is thus patchable.

In the example of the patchable phase singularity in Fig. \ref{fig:patchability}, consider the loops $g$ and $g'$ shown in panel b. Integrating $\dphi_{\Wn}$ over $g$ or $g'$ yields zero, since the wrapped phase first increases and then decreases along either loop. Indeed, the loop integral vanishes for any loop in $\Wn$, which is equivalent to patchability. Conversely, for the unpatchable absolute phase singularity, panel d shows a loop $g$ with $\int_{g'} \dphi_{\Wn} \neq 0$. Even though the integral vanishes over the other loop $g'$, the existence of a single loop with non-zero integral implies the absolute phase cannot be patched.

For a loop $g$ in $\Wn$, $\int_{g} \dphi_{\Wn} = 0$ if the loop can be continuously deformed to a single point in $\Wn$. The loops $g'$ in Fig. \ref{fig:patchability}b,d are such contractible loops. The integral vanishes because $\dphi_{\Wn}$ will be shown to be closed, i.e., its exterior derivative $\mathrm{d} \dphi_{\Wn} = 0$. Applying the exterior derivative $\mathrm{d}$ to a one-form generalizes the notion of taking a curl, as can be seen in Stokes' theorem \cite{weintraub14}
\begin{align}
\int_{g} \dphi_{\Wn} = \int_{G} \mathrm{d}\dphi_{\Wn}\,,\label{eq:stokes}
\end{align}
where $G$ is an oriented surface whose boundary is the loop $g$ with basepoint $\wb_1$. If we can find such a $G$, the right-hand side vanishes because $\mathrm{d}\dphi_{\Wn} = 0$, implying path-independence. Deforming $g$ continuously to $\wb_1$ -- essentially pulling $g$ tight like a lasso -- implicitly defines such a surface. Such a continuous deformation is called a homotopy \cite{fulton97}. If it exists, we say $g$ is homotopic to $\varepsilon_{\wb_1}$, the constant loop at $\wb_1$, with the basepoint $\wb_1$ held fixed. This is denoted by $g \simeq \varepsilon_{\wb_1}$. 

While it can be challenging to deal with loops in high-dimensional spaces with phase singularities and other holes, a clearer picture emerges in the complex plane.

\subsection{Loops in the punctured complex plane}
The integrals in $\Wn$ can also be evaluated in the complex plane using the coherence mapping $\gamma$, or $v$. Specifically, the punctured complex plane $\Wn$ that excludes the origin is used, as the phase singularities excluded from $\Wn$ get sent to the origin by $\gamma$. From \eqref{eq:phiugamma}, integration of $\dphi_{\Wn}$ along a loop $g$ in $\Wn$ is, through change of variables, equivalent to integrating $\dphi_{\Cn}$ along the loop $\gamma \circ g$ in $\Cn$, viz.
\begin{align}
\int_{g} \dphi_{\Wn} = \int_{\gamma \circ g} \dphi_{\Cn}\,. \label{eq:loopintegral}
\end{align}
Evaluation is facilitated by $\dphi_{\Cn}$ being closed \cite{weintraub14}, $\mathrm{d}\dphi_{\Cn}=0$. (From this, the earlier claim that $\dphi_{\Wn}$ is closed follows, as the pullback commutes with the exterior derivative.) The one-form $\dphi_{\Cn}$ in $\Cn$ is the canonical example of a closed one-form that is not exact, i.e., it cannot be written as the exterior derivative of a zero-form \cite{fulton97}. 

The integral $\int_{\gamma \circ g} \dphi_{\Cn}$ is determined by how often the loop $\gamma \circ g$ in $\Wn$ winds around the origin. The one-form $\dphi_{\Cn}$ measures the angle as seen from the origin \cite{fulton97}, yielding an integer multiple of $2 \pi$ for a loop $h$ in $\Cn$, viz.
\begin{align}
\int_{h} \dphi_{\Cn} = 2 \pi N(h)\,.\label{eq:winding}
\end{align}
The winding number $N(h) \in \mathbb{Z}$ counts the net number of turns $h$ makes around the origin and is positive for counterclockwise turns \cite{fulton97}. It follows that the unwrapped phases along two paths differ by an integer multiple of $2 \pi$, namely $N(\gamma \circ g)$ of the corresponding loop $g \in \Wn$.

Two loops in $\Cn$ have the same winding number if and only if they are homotopic to each other \cite{fulton97}, with the basepoint fixed. Consequently, a loop with winding number $N$ and basepoint $1 + \ju$ can be deformed into the circular loop $h_N(s) \equiv \cos(2\pi N s) +\ju \sin(2\pi N s)$ with $s \in [0, 1]$. Thus, integrating $\dphi_{\Cn}$ around a loop $h$ gives zero if and only if $N(h) = 0$, or equivalently, if and only if $h$ can be shrunk continuously to the constant loop $h_0$, i.e. $h \simeq h_0$. 

In the patchable example from Fig. \ref{fig:patchability}, the two loops $\gamma \circ g$ and $\gamma \circ g'$ are seen in the panel e) to both have a winding number of zero: neither encircles the origin, implying they are homotopic to each other and to the constant loop $h_0$. Conversely, the loop $\gamma \circ g$ in the unpatchable example in panel f) has a winding number of one, as it encircles the origin once in the positive direction.

\subsection{Identifying patchability in $\Wn$}
\label{sec:patchability}
The absolute phase singularities $A'$ are thus patchable if any arbitrary loop $g \in \Wn$ with basepoint $\wb_1$ gets mapped to a loop $\gamma \circ g$ in $\Cn$ with $N(\gamma \circ g) = 0$. No loop $g \in \Wn$ can have $N(\gamma \circ g) \neq 0$. This raises two questions, namely what characterizes loops $g$ in $\Wn$ with $N(\gamma \circ g) \neq 0$, and how can we tell whether such loops exist?

Two properties characterize a loop $g \in \Wn$ with $N(\gamma \circ g)) \neq 0$. First, it encircles a phase singularity (or other hole in $\Wn$) such that it cannot be continuously shrunk to the constant loop $\varepsilon_{\wb_1}$. If it were homotopic to $\varepsilon_{\wb_1}$, the integral would evaluate to zero due to \eqref{eq:stokes}, which in turn implies $N(\gamma \circ g) = 0$ due to \eqref{eq:loopintegral}, a contradiction. Second, it intersects $A'$, the set of absolute phase singularities on which the phase is defined. To see this, assume it did not, so that its support is in $\Wa$. However, $\dphi_{\Wn} = \mathrm{d}\phia$ and thus exact in $\Wa$, implying $N(\gamma \circ g)$ = 0, yielding a contradiction. Both properties apply to $g$ in Fig. \ref{fig:patchability}b.

%Because $g$ cannot be the constant loop, there exists a point $\wb_2 \neq \wb_1$ in its support. For this $\wb_2$, the phase unwrapping integral along the absolute-phase path $\Gamma^{\mathrm{a}}_{\wb_2}$ from $\wb_1$ to $\wb_2$ differs from the integral along the alternate path from $\wb_1$ to $\wb_2$ obtained by concatenating $g$ with the absolute-phase path. Specifically,
%\begin{align}
%\int_{\Gamma^{\mathrm{a}}_{\wb_2}} \dphi_{\Wn} - \int_{g \cdot \Gamma^{\mathrm{a}}_{\wb_2}} \dphi_{\Wn} = - 2 \pi N(\gamma \circ g) \neq 0\,,\label{eq:pdepWn}
%\end{align}
%because the two $\Gamma^{\mathrm{a}}$ contributions cancel. By assumption, the support of these two paths is within $\Wa$. However, $\dphi_{\Wn} = \mathrm{d}\phia$ and thus exact in $\Wa$, implying that the left-hand side evaluates to zero, yielding a contradiction. It follows that $g$ intersects $A'$.

To test whether a loop $g$ with $N(\gamma \circ g) \neq 0$ exists, we turn to the fundamental group of $\Wn$ \cite{hatcher01}. The fundamental group $\pi_1(\Wn, \wb_1)$ of $\Wn$ comprises all homotopy equivalence classes of loops with basepoint $\wb_1$. The equivalence class represented by the loop $g$ is denoted by $[g]$, and a loop $g'$ is in this class, $g' \in [g]$, if $g$ and $g'$ are homotopic, i.e. $g \simeq g'$. The group operation is concatenation, and $\varepsilon_{\wb_1}$ is the identity element. No loop $g$ with $N(\gamma \circ g) \neq 0$ exists if and only if $N(\gamma \circ h) = 0$ for all $[h]$ in the fundamental group $\pi_1(\Wn, \wb_1)$. If $H$ is a set of generators in a presentation of the fundamental group, the test can be restricted to all $[h] \in H$. Another way to interpret this test is to require that $\mathrm{im}(\gamma_{*}) = \{ h_0\}$, i.e., that the image of the induced homomorphism of $\gamma$, $\gamma_{*}$, only comprise $h_0$, the constant loop in $\Cn$. The induced homomorphism thus needs to map $\pi_1(\Wn, \wb_1)$ to the trivial group (cf. Proposition 1.33 in \cite{hatcher01} for a covering space perspective). While identifying $\pi_1(\Wn, \wb_1)$ is generally difficult, the test is easy to apply. 

\section{Reconstructing the absolute phase}
\label{sec:reconstruct}
We turn to reconstructing the absolute phase from InSAR observations. The reconstructed absolute phase may serve as input to displacement or elevation retrievals, but the reconstruction does not address whether the absolute phase permits such geometric interpretation as a range difference. 

Under what conditions can the absolute phase be reconstructed from discrete InSAR measurements? A general two-step approach \cite{zebker98,rosen00} for a multidimensional interferogram is to i) unwrap the phase to obtain the absolute phase up to a uniform offset, an integer multiple of $2 \pi$; and ii) estimate this offset using such strategies as multifrequency observations.

\subsection{Multidimensional interferograms}
\label{sec:multidimensional}
In a multidimensional interferogram, the absolute phase may be reconstructed through the slow variation of the phase along these dimensions, most commonly space or time \cite{ghiglia98}. In an ideal scenario, phase unwrapping reconstructs the absolute phase up to an integer multiple of $2 \pi$, the estimation of which is deferred to Sec. \ref{sec:isolated}. For now, the focus is on the conditions under which the absolute phase can be reconstructed up to a single constant by unwrapping, treating idealized continuous and then discretely sampled observations.

\subsubsection{Continuous unwrapping amid singularities}
\label{sec:continuous}
\paragraph{Mathematical notation}
The complex coherence $\gamma$ is assumed to vary smoothly in the space $X \cong S \subset \mathbb{R}^n$. For instance, $n = 2$ for a spatial interferogram. Discontinuities in $\gamma$ (e.g., layover) have been excised from $X$. Let the set $\Xn = X \setminus \gamma^{-1}\left[\{0\}\right]$, obtained by removing phase singularities, be open, path connected and locally path connected \cite{hatcher01}. The wrapped phase $\varphi = \mathrm{arg} \gamma$ is assumed to arise congruently from a smooth absolute phase field $\phia: \Xa \rightarrow \mathbb{R}$, where $\Xa = X \setminus A$, the space $X$ excluding all absolute phase singularities $A$. For instance, in differential InSAR with given observation times $t_1$ and $t_2$, a smooth $\phia(\xb)$ over $X$ is obtained whenever the unobserved coherence $\gamma(\xb, t)$ at $t \in [t_1, t_2]$ with fixed primary $t_1$ is a smooth function of space $\xb$ and time $t$; however, vanishing coherence at $\xb$ and some intermediate $t$ induces an absolute phase singularity at $\xb$ at secondary time $t_2$.

\paragraph{Path-based unwrapping}
The phase unwrapping considered here consists of integrating $\dphi_{\Xn}$ along a path. The differential one-form $\dphi_{\Xn}$ is defined in analogy to Sec. \ref{sec:unwphasedef} and also closed. Given an observed $\gamma(\xb)$ on $\Xn$, an unwrapped phase difference between points $\xb$ and $\xb'$ can be computed through $\Delta \phi(\Gamma) = \int_{\Gamma} \dphi_{\Xn}$ along a path $\Gamma: [0, 1] \rightarrow \Xn$ with $\Gamma(0) = \xb$ and $\Gamma(1) = \xb'$. For fixed endpoints $\xb$ and $\xb'$, $\Delta \phi$ may depend on the path. 

\paragraph{Absolute phase reconstruction}
The absolute phase can be reconstructed up to a uniform offset if all unwrapping paths are restricted to $\Xa$. Reconstructability in a path-connected space $\Xa$ follows because the observed differential form $\dphi_{\Xn}$ is identical to $\mathrm{d}\phia$ on $\Xa$ due to the assumed smoothness and congruence. The goal, then, is to avoid absolute phase singularities during unwrapping.

\begin{figure}
		\centering
			\includegraphics{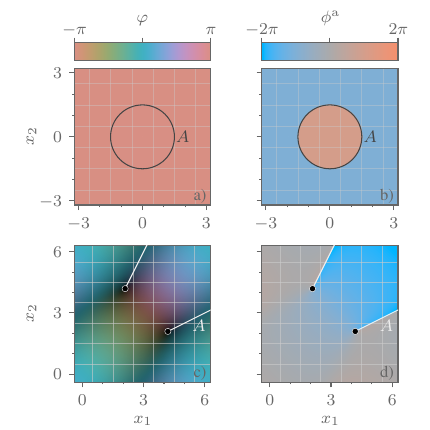}
		\caption{Continuous, noise-free two-dimensional wrapped phase $\varphi$ (left, with darker colors corresponding to lower coherence) and absolute phase (right) modeled for two differential InSAR scenarios. Scenario 1 (a--b) consists of two countermoving targets, while scenario 2 (c--d) adapts the three surface segments from Fig. \ref{fig:threesurf}. In both, the scatterer positions vary with $x_1$ and $x_2$. \label{fig:unwrap}}
\end{figure}

Fig. \ref{fig:unwrap} illustrates two challenges to absolute phase reconstruction through unwrapping. The absolute phase singularities cannot be observed and --- in contrast to the integrals in the configuration space $\Wn$ --- need not bear any relationship with phase singularities or holes in the domain. The two-dimensional differential InSAR interferogram in Fig. \ref{fig:unwrap}a) has a uniform wrapped phase of $\pi$, perfect coherence, and no path dependence. Nevertheless, it contains a circular absolute phase singularity $A$, where the coherence vanished at some intermediate time. Specifically, this interferogram was generated from \eqref{eq:decorrelation} with spatially variable $f(s) = 2/3 \sqrt{x_1^2 + x_2^2}$. Similar to Fig. \ref{fig:decorrelation}, the points inside and outside $A$ passed the origin on opposite sides, which manifests in the unobserved absolute phase of Fig. \ref{fig:unwrap}b but not in the observed interferogram. The circular absolute phase singularity disconnects $\Xa$, imperiling reconstruction by unwrapping even if the location of $A$ were known. Such disconnection does not occur in Fig. \ref{fig:unwrap}c--d, obtained from the three surface segments model (Fig. \ref{fig:threesurf}) with final surface positions $b_1$ and $b_2$ varying in space like $x_1$ and $x_2$, respectively. Absolute phase singularities radiate outward from the phase singularities but do not connect them. 

Are absolute phase singularities related to branch cuts? Branch cuts are barriers erected during phase unwrapping to circumvent path dependence \cite{goldstein88,rosen00}. For instance, in Fig. \ref{fig:unwrap}c, the two phase singularities could be connected by a branch cut, which enables recovery of an unwrapped phase field outside the branch cut. However, this unwrapped phase bears no resemblance to the absolute phase in Fig. \ref{fig:unwrap}d. To recover the absolute phase, branch cuts should be placed at absolute phase singularities. Doing so can be difficult, as absolute phase singularities need not have observable manifestations (Fig. \ref{fig:unwrap}a). However, the strategy of linking the two phase singularities with a branch cut is appropriate when path dependence is induced by localized noise, to be studied in discrete data.

\subsubsection{Discrete unwrapping amid singularities}
\label{sec:discrete}
\paragraph{Mathematical notation}
Consider a discretely sampled complex coherence $\gamma_{\mathrm{d}}: X_{\mathrm{d}} \rightarrow \mathbb{C}$ defined on $X_{\mathrm{d}} \subset \mathbb{Z}^n$, a regular lattice with sample points spaced equidistantly along the coordinate directions. The coherence $\gamma_d$ is assumed to be obtained by sampling a smooth coherence field $\gamma$ that arises congruently from a smooth absolute phase $\phia$, as in Sec. \ref{sec:continuous}. Generically, $\phia$ will be defined on all $X_{\mathrm{d}}$.

\paragraph{Discrete unwrapping}
Path-based phase unwrapping proceeds along coordinate-parallel segments connecting adjacent lattice points. A space $X_{\mathrm{d}}\subset \mathbb{Z}^n$ is called lattice path connected if any pair of points can be connected by such a path traversing adjacent pixels. Phase unwrapping along a unit step in $x_i$, denoted by $\Delta_i$, evaluates
\begin{align}
\Delta \phi_{\mathrm{d}}(\xb, \xb+\Delta_i)= \arg \left(\gamma_{\mathrm{d}}(\xb+\Delta_i) \gamma_{\mathrm{d}}(\xb)^{\star}\right)\,, \label{eq:discreteunwrapping}
\end{align}
which is in the interval $(-\pi, \pi]$. 

\paragraph{Reconstructability}
Absolute phase reconstruction by unwrapping aims to recover $\phia_{\mathrm{d}}(\xb)$ on $X_{\mathrm{d}}$. By this we mean that the reconstruction $\hat{\phia}_{\mathrm{d}}$ matches the true absolute phase $\phia_{\mathrm{d}}$ for all $\xb \in X_{\mathrm{d}}$ up to a uniform $2 \pi n$. 

The absolute phase can be reconstructed on a path-connected $X_{\mathrm{d}}$ up to an offset by unwrapping if unwrapping along each segment is correct, i.e., if
\begin{align}
\Delta \phi_{\mathrm{d}}(\xb, \xb+\Delta_i) = \phia_{\mathrm{d}}(\xb+\Delta_i) - \phia_{\mathrm{d}}(\xb)
\end{align}
for all $i$ and $\xb$. As $\Delta \phi_{\mathrm{d}}(\xb, \xb+\Delta_i) \in (-\pi, \pi]$, it follows that 
\begin{align}
\left|\phia_{\mathrm{d}}(\xb+\Delta_i) - \phia_{\mathrm{d}}(\xb)\right| < \pi \,, \label{eq:itoh}
\end{align}
known as Itoh's criterion \cite{itoh82}. Itoh's criterion requires the absolute phase to change slowly between adjacent lattice points and can be violated due to steep gradients (undersampling) or apparent discontinuities \cite{zebker98,pepin24}. The required slow variation warrants re-interpretation amid absolute phase singularities.

Itoh's criterion can be violated due to absolute phase singularities, in addition to rapid phase changes or noise. An example is shown in Fig. \ref{fig:unwrap}a. Assuming the continuous coherence $\gamma(\xb)$ shown in the figure is sampled at integer lattice points to yield $\gamma_{\mathrm{d}}(\xb)$, the wrapped phase is $\pi$ throughout, but unwrapping across the absolute phase singularity surrounding the point at $x_1 = x_2 = 0$ cannot pick up the $2 \pi$ jump in absolute phase (Fig. \ref{fig:unwrap}b). Thus, it is not sufficient for the coherence and wrapped phase to change slowly. Across the absolute phase singularity, reconstruction breaks down because Itoh's criterion is violated. The violation is analogous to a discontinuous displacement or elevation difference, but the discontinuity is always an integer multiple of $2 \pi$.

\begin{figure}
		\centering
			\includegraphics{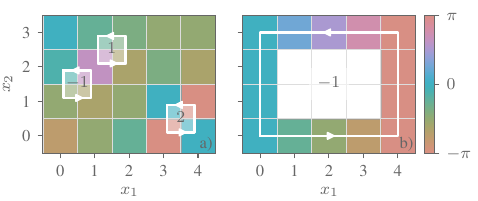}
		\caption{Loops identify path dependence in discrete, two-dimensional interferograms through a non-zero winding number $N$, printed inside a loop. a) Elementary loops are shown around two generic phase singularities with $N = \pm 1$ and a non-generic singularity with $N = 2$. b) The large loop with $N = 1$ encloses a hole rather than a phase singularity. \label{fig:residue}}
\end{figure}

\paragraph{Phase and absolute phase singularities}
Certain violations of Itoh's criterion can be detected through path dependence of phase unwrapping \cite{goldstein88}. A nonzero unwrapped phase along a loop identifies a violation. As in the continuous case, these nonzero unwrapped phases occur as integer multiples of $2 \pi$. Elementary 2 $\times$ 2 pixel loops (see Fig. \ref{fig:residue}a for two dimensions) can yield nonzero $N$.

An elementary loop with nonzero $N$ is said to encircle a phase singularity, by analogy to the continuous case. In InSAR, such a loop and $N$ are referred to as residue \cite{bamler98,rosen00}. In two dimensions, a standard loop orientation can be chosen \cite{berry09}, assigning generic phase singularities a value of $N = 1$ (positive) or $N = -1$ (negative). The elementary loop integral is a discrete exterior derivative operation such as the curl \cite{goldstein88,bamler98}. In this discrete setting, the only non-generic singularity has $N = 2$. This non-generic singularity is shown in Fig. \ref{fig:residue}a): Each elementary loop segment contributes $\Delta \phi_{\mathrm{d}} = \pi$, due to the restriction to $(-\pi, \pi]$. The same result would be obtained for the inverse loop because the unwrapped phase is restricted to $(-\pi, \pi]$. This violation of antisymmetry, $\pi = \Delta \phi_{\mathrm{d}}(\xb, \xb+\Delta_i) \neq - \Delta \phi_{\mathrm{d}}(\xb+\Delta_i, \xb)$, challenges discrete approximations of the exterior derivative \cite{goldstein88,bamler98}. These complications are usually neglected in practice because they do not arise generically.

If $X_{\mathrm{d}}$ contains no holes, elementary loops suffice to detect path dependence \cite{costantini98}. If $X_{\mathrm{d}}$ contains holes covering lattice points, path dependence is detected through loops encircling these holes (Fig. \ref{fig:residue}b). Such holes are commonly encountered in practice, for instance when water bodies are masked.

Itoh violations due to absolute phase singularities in $X_{\mathrm{d}}$ cannot be detected using path dependence. As in the continuous case (Sec. \ref{sec:continuous}), unpatchable absolute phase singularities are not directly visible in $X_{\mathrm{d}}$. The discrete unwrapping in \eqref{eq:discreteunwrapping} is insensitive to phase jumps by integer multiples of $2 \pi$ across these absolute phase singularities.

\paragraph{Branch cuts for absolute phase reconstruction}
Branch cuts should cover absolute phase singularities, but branch cut placement strategies are geared toward erasing path dependence in the interferogram. These strategies generally connect positive and negative phase singularities. Let us quickly review the rationale, accounting for noise. Consider a phase field meeting \eqref{eq:itoh} and apply a localized, noise-like perturbation in a simply connected interior subset of $X_{\mathrm{d}}$. As an example, the pixel at $(x_1, x_2) = (1, 2)$ in Fig. \ref{fig:residue}a) may be thought of as having been perturbed, which induced path dependence. In two dimensions, such localized path dependence implies residues whose total charge is zero because loops surrounding the perturbation remain unaffected \cite{freund95}. The pair of residues with $N = \pm 1$ in Fig. \ref{fig:residue}a) has vanishing total charge. Such an oppositely charged pair arises generically as the perturbation magnitude increases: the two residues nucleate together via bifurcation and then separate \cite{freund95,adachi07}. Phase unwrapping errors can be contained by branch cuts connecting such a residue pair, or, more generally, residues with vanishing total charge \cite{goldstein88}. 

Branch cut strategies connecting residue pairs fail for unpatchable absolute phase singularities because, ultimately, absolute phase singularities generally bear no relation with residues. The are invisible in the discrete wrapped phase (Fig. \ref{fig:unwrap}) and do not manifest through path dependence in the interferogram. Similar challenges exist in higher dimensions. In three dimensions, localized disturbances induce path dependence around phase singularities that are generically loop shaped \cite{huntley01,adachi07} and could be contained by a branch-cut membrane spanning the loop. In summary, absolute phase singularities should be covered by branch cuts, but the placement of branch cuts based on the observed interferogram is challenging.

\subsection{Integer offset estimation}
\label{sec:isolated}
\subsubsection{Single coherence observation}
Two strategies for reconstructing the absolute phase from a single isolated coherence, in turn derived from one primary and one secondary observation, are commonly considered \cite{zebker98,rosen00}. In practice, they are commonly applied to multidimensional observations through averaging or selection of individual pixels.

\paragraph{Ancillary information}
This strategy is viable whenever the absolute phase $\phia$ is known to be defined for the given observation, thus ruling out an absolute phase singularity \cite{zebker98}. If it is known to be small in magnitude, $|\phia| < \pi$, the absolute phase $\phia = \varphi$, where the wrapped phase $\varphi$ is taken to be in the interval $(-\pi, \pi)$. Unwrapping along the (unobserved) absolute phase path $\Gamma^{\mathrm{a}}$ is unnecessary. More generally, the absolute phase may be reconstructed if it can be externally constrained to better than $2 \pi$. Such a constraint will generally be easier to apply if path dependence can be ruled out. 

\paragraph{Multifrequency observations}
This strategy applies to point-like targets that can be described by \eqref{eq:distance}, for which absolute phase singularities cannot occur. The dependence of the absolute phase on the difference in range (or optical path) in \eqref{eq:distance} is proportional to the wavenumber $k$ and thus frequency, allowing for the estimation of the range difference within a wider ambiguity band than the first strategy \cite{madsen93}. This method, along with radargrammetry, is used in across-track interferometry, usually by averaging over multiple observations due to the high noise level \cite{brcic09}. However, \eqref{eq:distance} breaks down when the vertical extent approaches the height of ambiguity or, in differential InSAR, when the displacement variability is nonnegligible compared to the wavelength, or for dielectric changes \cite{bamler98,dezan15,zwieback16}. Ionospheric phase contributions also invalidate \eqref{eq:distance}. Three frequencies facilitate estimating the absolute ionospherically corrected phase and detecting deviations from point-target behavior.

\subsubsection{Samples along the absolute path}
Serial observations along the absolute path facilitate absolute phase reconstruction. Assume that the ``absolute path dimension'' is not included in the space $X_{\mathrm{d}}$. For instance, in the BorealScat data (Sec. \ref{sec:borealscat}), $X_{\mathrm{d}}$ encodes a spatial interferogram while the absolute path traverses time, so that integer phase offset estimation can be applied independently for each spatial pixel $x \in X_{\mathrm{d}}$. The discrete samples along the absolute path provide a sampled approximation to \eqref{eq:phia} is available. The denser the sample spacing, the easier the unwrapping and detection of zero or near-zero coherence magnitudes. In the BorealScat example, the 5-min spacing between successive secondary acquisitions facilitated reconstruction of \eqref{eq:phia} for the 1-day interferogram. However, determining unambiguously whether nearby points A and B in Fig. \ref{fig:borealscat} differed in absolute phase by $\sim 2 \pi$ would require infinitesimally dense samples. In across-track interferometry, multiple samples along the absolute path correspond to multiple baselines, the analysis of which resembles multifrequency processing for point-like targets \cite{lombardini98}.

For sparsely sampled observations along the absolute path, generalized closure phases can provide contextual information on absolute phase singularities \cite{dezan15,zwieback24}. The simplest case is the closure phase \cite{dezan15}
\begin{align}
\Xi(s_0, s', s_1) = \arg\left(\gamma(s'; s_0) \gamma(s_1; s') \gamma^*(s_1; s_0)\right)
\end{align}
formed from three acquisitions at three positions along the absolute path, $s_0 \leq s' \leq s_1$. It depends on coherences formed from different primary acquisitions at $s_0$ and $s'$ (the second argument of $\gamma$), in contrast to the absolute phase $\phia(s_1; s_0)$ with primary $s_0$. This added complexity makes it difficult to constrain absolute phase reconstruction. However, large closure phase magnitudes indicate a loss of correspondence between phase and range difference, as occurs for non-point-targets such as permafrost landforms with nonuniform displacement (Fig. \ref{fig:threesurfts}) or vegetation \cite{dezan15,zwieback16,ansari21}. The insensitivity of closure phases to atmospheric phase screens makes them amenable to operational observational analyses.

\section{Conclusion}
The proposed general definition \eqref{eq:phia} of the absolute phase is purely observational and universally applicable. It is consistent with and extends previous definitions for point-like targets, whose absolute phase is proportional to the range difference under simplifying assumptions. 

This general absolute phase exhibits challenging properties. Its direct observation would require continuous measurements along the absolute path, while a regular interferogram only samples the endpoints. Fig. \ref{fig:threesurfts} illustrates the need for continuous observations: the absolute phase is path dependent, determined by intermediate states between $s = 0$ and $s = 1$ that discrete observations cannot capture. Path dependence implies that the absolute phase is not proportional to a range difference. Sec. \ref{sec:patch} shows that path dependence in the configuration space $\Wn$ is equivalent to the existence of unpatchable absolute phase singularities, with absolute phase jumps by nonzero integer multiples of $2 \pi$. In observations, these absolute phase singularities can occur even for perfect coherence (Fig. \ref{fig:unwrap}), and they complicate observation-based reconstruction of the absolute phase.

The absolute phase usually needs to be reconstructed using assumptions. For multidimensional interferograms, multifrequency methods (for point-like targets) or external information are commonly combined with phase unwrapping. Phase unwrapping can reconstruct the absolute phase up to a constant, provided certain conditions are met. Absolute phase singularities are a challenge as they generally leave no trace in the interferogram, and yet unwrapping paths are to avoid them. Sec. \ref{sec:multidimensional} re-interprets conditions in which the absolute phase can be reconstructed up to a constant through unwrapping \cite{itoh82}, accounting for absolute phase singularities. The reconstructed absolute phase may serve as input to displacement or elevation retrieval algorithms, though this assumes proportionality to range difference.

How useful is the absolute phase as defined here, given these challenges? Previously, the absolute phase formed the basis for displacement and elevation retrieval \cite{madsen93,rosen00}. This relied on the assumption that it is proportional to range difference, an assumption valid for point targets. In general, however, this correspondence cannot hold. In differential interferometry the observational definition is the only way to assign a real quantity that is congruent with the wrapped phase and continuous (Sec. \ref{sec:differential}). The loss of correspondence to range difference severs the absolute phase's direct connection to the equivalent displacement or phase center position practitioners seek to estimate. Downstream algorithms need to test whether the estimated absolute phase corresponds to a range difference, for instance through closure phases. Although the absolute phase need not have a simple physical interpretation, the observational definition and the theoretical tools introduced here enable principled evaluation of InSAR processing chains.

\bibliographystyle{ieeetr}%ieeetr
\bibliography{absphasebib}

\end{document}